\documentclass[pdftex,citeautoscript,aps,prl,superscriptaddress,cite,amsmath,amsfonts,amssymb,twocolumn]{revtex4}
\pdfoutput=1
\renewcommand{\vec}[1]{\mathbf{#1}} 
\usepackage{graphicx}
\usepackage{color}
\usepackage{url}

\renewcommand{\Re}{\operatorname{Re}}
\renewcommand{\Im}{\operatorname{Im}}

\newcommand{\figref}[1]{Fig.~\ref{fig:#1}}

\newcommand{\Figref}[1]{Figure~\ref{fig:#1}}

\newcommand{\eqnumref}[1]{(\ref{eq:#1})}
\renewcommand{\eqref}[1]{Eq.~\eqnumref{#1}}

\newcommand{\citeasnoun}[1]{Ref.~\onlinecite{#1}}

\begin{document}
\title{Modeling near-field radiative heat transfer from sharp objects using a general 3d numerical scattering technique}
\author{Alexander~P.~McCauley}
\affiliation{Department of Physics, Massachusetts Institute of Technology, Cambridge MA 02139, USA}
\author{M.~T.~Homer~Reid}
\affiliation{Department of Physics, Massachusetts Institute of Technology, Cambridge MA 02139, USA}
\affiliation{Research Laboratory of Electronics, Massachusetts Institute of Technology, Cambridge MA 02139, USA}
\author{Matthias~Kr\"{u}ger}
\affiliation{Department of Physics, Massachusetts Institute of Technology, Cambridge MA 02139, USA}
\author{Steven~G.~Johnson}
\affiliation{Department of Mathematics, Massachusetts Institute of Technology, Cambridge MA 02139, USA}

\begin{abstract}
We examine the non-equilibrium radiative heat transfer between a plate
and finite cylinders and cones, making the first accurate theoretical
predictions for the total heat transfer and the spatial heat flux
profile for three-dimensional compact objects including corners or
tips.  We find qualitatively different scaling laws for conical shapes
at small separations, and in contrast to a flat/slightly-curved
object, a sharp cone exhibits a local \emph{minimum} in the spatially
resolved heat flux directly below the tip.  The method we develop, in
which a scattering-theory formulation of thermal transfer is combined
with a boundary-element method for computing scattering matrices, can
be applied to three-dimensional objects of arbitrary shape.
\end{abstract}
\maketitle

\begin{figure}[tb]
\includegraphics[width=0.45\textwidth]{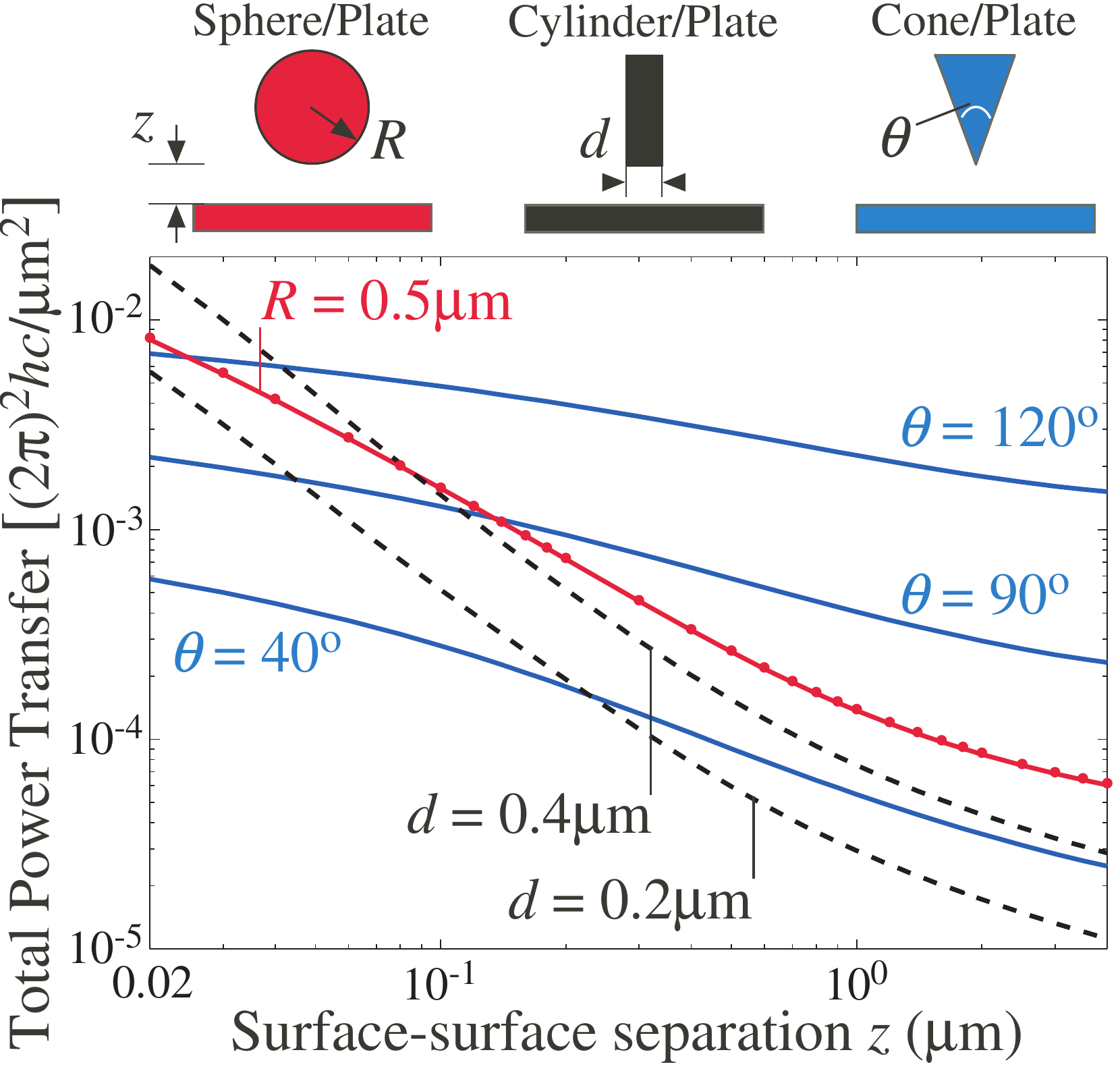}
\centering
\caption{Total thermal transfer between a silica plate and doped
  silicon objects of various shapes.  The plate is semi-infinite, and
  the objects all have height equal to $1\,\mu\mathrm{m}$ in the
  $z$-direction.  The plate/environment temperature is $T_P =
  300\,\mathrm{K}$ and the objects are at temperature $T_A
  =600\,\mathrm{K}$.  Red dots denote results with the sphere
  scattering matrix determined analytically, the only case in which we
  have an analytic solution for the scattering matrix.}
\label{fig:total_power}
\end{figure}

\emph{Introduction}: We make the first accurate theoretical
predictions for near-field thermal transfer from 3d compact objects of
arbitrary shape (including corners or tips) to a dielectric substrate.
Our work is motivated by studies of non-contact thermal writing with a
hot, sharp object~\cite{Mamin96, Wilder98}.  Theory has
predicted~\cite{PolderVanHove71, Volokitin01:PRB} and experiments have
confirmed~\cite{Shen09, RosseasuSiria09} that radiative heat transfer
between two bodies at different temperatures is greatly enhanced as
their separation is reduced to sub-micron scales, due to contributions
from evanescent waves.  Until the last few years, the only rigorous
theoretical results for thermal transfer concerned parallel plates;
however, very recently rigorous theoretical predictions for
sphere-sphere~\cite{Narayanaswamy08:spheres} and
sphere-plate~\cite{Kruger11, Otey11} geometries as well as general
formalisms for planar structures~\cite{bimonte09} and arbitrary
shapes~\cite{Messina11, Kruger11} have been presented.  Nevertheless,
such techniques were previously implemented only when analytic
expressions for the scattering matrices were known (e.g., spheres and
plates in 3d).  As an alternative, stochastic finite-difference
time-domain methods have been used to examine heat transfer for
periodic structures~\cite{Rodriguez11:ThermalFDTD}, but this method is
not computationally well-suited for compact objects in three
dimensions.  Our technique extends the formalism
of~\citeasnoun{Kruger11} directly to arbitrary compact objects.  To do
this, we use a boundary-element method in which the object is
described by a generic surface mesh~\cite{RWG82}.  We then numerically
compute the scattering matrices of this object in a multipole basis;
for our study, we employ a cylindrical-wave basis.  Unlike the usual
spherical-wave basis, this allows us to concentrate our resolution on
the surfaces adjacent to the substrate, but requires a new quadrature
approach to discretize the scattering matrix.  In addition to
sphere-plate heat transfer, we study both cylinder-plate and
cone-plate configurations (see sketch in~\figref{total_power}), for
which no known analytic solution exists.  Our results exhibit clear
scaling laws for the total heat transfer that distinguish locally flat
structures (e.g., cylinders and spheres), from locally sharp
structures (cones).  In addition, we study the spatial distribution of
heat-flux over the substrate, a topic that has been treated previously
using a point-dipole approximation for the heat
source~\cite{Mulet2001}.  Our results show that the heat flux pattern
depends strongly on the shape of the tip.  Cones in particular have a
flux pattern exhibiting an unusual feature: a local \emph{minimum} in
the heat flux directly below the tip, which we can explain with a
modified dipole picture.



\emph{Method}: In our setup, an object $A$ at (local) temperature
$T_A$ faces a dielectric plate $P$ at temperature $T_P$, in an
environment $E$ that is also at temperature $T_P$.  We use the
framework of Rytov's theory~\cite{Rytov89_3}, in which all sources
emit radiation independently.  The full non-equilibrium Poynting flux
can be computed with radiative sources from $P$ and $E$ only, as the
flux from $A$ at temperature $T_A$ must equal the flux from $P$ and
$E$ at temperature $T_A$ (with opposite sign), due to detailed
balance~\cite{Eckhardt84}.  To compute the power flux, we first
compute the non-equilibrium electric field correlator $ \langle
\vec{E}(\vec{x}) \otimes \vec{E}^*(\vec{x}^\prime)\rangle_j$ due to
radiation from $j=P, E$ for general $\vec{x}\neq\vec{x}^\prime$ (the
Poynting flux will be obtained at the end by taking
$\mathrm{lim}_{x^\prime \rightarrow x} \nabla_{x^\prime} \langle
\vec{E}\otimes \vec{E}^{\prime *}\rangle$) which is expressed as an
integral of the general form:
\begin{equation*}
\langle \vec{E}\otimes \vec{E}^{\prime *}\rangle_j
=
\int_0^\infty \frac{d\omega}{2\pi} \Theta(\omega, T_j)
\langle\vec{E}\otimes\vec{E}^{\prime *}\rangle_{j,\omega},
\end{equation*}
where $\Theta = \omega^4 \left(\exp\left(\hbar \omega / k_B T\right) -
1\right)^{-1}$~\cite{Rytov89_3, Eckhardt84}, $\hbar$ is Planck's
constant and $k_B$ the Boltzmann constant.  Unless otherwise noted, we
consider each frequency $\omega$ separately and drop the $\omega$
subscript below.

The correlator takes on a simple form in an orthogonal basis
$\vec{E}_\alpha(\omega; \vec{x})$ for the field degrees of freedom (in
our case, these will be cylindrical waves in the $\pm z$ direction),
indexed by a (discrete or continuous) index $\alpha$, and represent
the correlator as a matrix $\mathbb{D}$.  In matrix notation (with
implied summation over repeated indices):
\begin{equation}
\langle \vec{E}(\vec{x})\otimes \vec{E}^*(\vec{x}^\prime)\rangle =
\left(\mathbb{D}\right)_{\alpha^\prime, \alpha}
\vec{E}_{\alpha^\prime} (\vec{x}) \otimes \vec{E}^*_\alpha (\vec{x}^\prime).
\label{eq:FreeCorrelator}
\end{equation}
$\mathbb{D} = \mathbb{D}_P + \mathbb{D}_E$ due to statistical
independence of the thermal fluctuations, where $\mathbb{D}_{P/E}$
involve sources only from $P$ / $E$.  The correlators
$\mathbb{D}_{P/E}$ are obtained from the ``unperturbed'' correlators
$\mathbb{D}^0_{P/E}$; $\mathbb{D}^0_P$ involves the plate sources
without $A$ and $\mathbb{D}^0_E$ involves the environment sources with
neither $A$ nor $P$ present.  The $\mathbb{D}^0_j$ are known
analytically (see below), and the full correlators $\mathbb{D}_j$ can
be determined from them by use of the Lippmann-Schwinger
equation~\cite{Kruger11, Rahi09:PRD}.  In our notation:
\begin{eqnarray}
\mathbb{D}_{j} &=& \mathbb{O}_{j} \mathbb{D}_{j}^0 \mathbb{O}_{j}^\dagger, ~~~j = P, E 
\end{eqnarray}
The $\mathbb{O}_j$ are matrices that describe the scattering of
incoming and outgoing fields with the allowance for sources in between
the objects, described explicitly in~\cite{Kruger11:long}.  These are
constructed from the more conventional incoming/outgoing scattering
matrices $\mathbb{F}_{P/A}$~\cite{Merzbacher98, Rahi09:PRD} for
objects $P$ and $A$ individually.  As object $P$ is a plate,
$\mathbb{F}_P$ is known analytically.  However, $\mathbb{F}_A$ cannot
be determined analytically for a general object $A$.  Instead, the
computation of the scattering matrix elements is accomplished via a
boundary-element method~\cite{RWG82}, described below.  The
$z$-component of the Poynting flux at position $\vec{x}$, $S_\vec{x}$,
and the total power flux $S_T$ through the $z = 0$ plane can both be
expressed as operator traces: $S_{\vec{x} / T} = \Re \mathrm{Tr}\left[
  \mathbb{S}_{\vec{x} / T} \mathbb{D}_T\right]$, with
$\left(\mathbb{S}_\vec{x}\right)_{\alpha^\prime, \alpha} =
-\frac{i}{\omega} \hat{\vec{z}}\cdot
\left[\vec{E}_{\alpha}(\vec{x})\times \left(\nabla \times
  \vec{E}_{\alpha^\prime}(\vec{x})\right)^*\right]$ and
$\left(\mathbb{S}_T\right)_{\alpha^\prime, \alpha}$ given below.


We employ a cylindrical-wave basis of fields
$\vec{E}_{s,m,k_\rho,p}(\vec{x})$ in which the waves (also known as
Bessel beams) propagate in the $\pm z$ direction~\cite{Tsang2000}.
The variable $s=\pm$ refers to the direction of propagation; $m$ is
the (integer) angular moment of the field, $0 \leq k_\rho < \infty$
the radial wavevector, and $p = M, N$ the polarization.  The composite
index in this case is $\alpha = \lbrace s,m,k_\rho, p\rbrace$.  This
basis is especially well-suited to the case considered here in which
objects have rotational symmetry about the $z$-axis, as different
values of $m$ are decoupled.  

To compute the elements of $\mathbb{F}_A$, we use a boundary-element
method (BEM)~\cite{RWG82, Reid10:thesis}.  In this framework, the
surface of object $A$ is discretized into a mesh; our numerical method
then computes the induced currents from an incident multipole field
$\vec{E}_\alpha(\vec{x})$ (here $\alpha = \lbrace
s,m,k_\rho,p\rbrace$).  The multipole moments of this current
distribution are then computed in a straightforward
manner~\cite{Tsang2000}, which yield the scattering matrix
$\mathbb{F}_A$~\cite{Rahi09:PRD}.  Because the cylindrical-wave basis
distinguishes between waves in the $\pm z$ direction (unlike a
spherical wave basis), and because the near-field thermal transport
mostly depends on reflections from adjacent surfaces, we are able to
concentrate most of our BEM mesh resolution on the part of the surface
of $A$ nearby the plate, greatly improving computational efficiency.
For example, in the mesh for a cone below we use $\sim 250$ times more
resolution at the tip than at the base.

One complication of cylindrical multipoles is that $k_\rho$ is a
continuous index and matrix multiplication is turned to integration.
For computational purposes, this integration must be approximated as a
discrete sum by numerical quadrature. We approximate the integral over
$k_\rho$ using a Gaussian quadrature scheme~\cite{AbramowitzStegun72}
for high accuracy.  For example, consider the scattering matrix
$\mathbb{F}_A$ of object $A$; its action on an incident electric field
can be discretized as (for simplicity, summation over $m$ and $p$ is
suppressed): $\mathbb{F}_A \vec{E}_{k_{\rho,i}} = \int_0^\infty
\frac{dk_\rho^\prime}{2\pi}
\left(\mathbb{F}_A\right)_{k_\rho^\prime;k_{\rho, i}}
\vec{E}_{k_\rho^\prime} \approx \sum_{j=0}^N w_j
\left(\mathbb{F}_A\right)_{j,i} \vec{E}_{k_{\rho,j}}$ where the sets
$\lbrace w_j, k_{\rho,j} \rbrace$ form a set of one-dimensional
quadrature weights and points, respectively, and $\left(
\mathbb{F}_A\right)_{j,i} =
\left(\mathbb{F}_A\right)_{k_{\rho,j},k_{\rho,i}}$ are the elements of
the continuous scattering matrix.

The analytic expression for the non-equilibrium electric field
correlator of a plate at temperature $T_P$ and environment at $T = 0$
expressed in the planewave basis is well-known~\cite{Rytov89_3,
  Kruger11:long}.  Since there is a standard identity relating
planewaves to cylindrical waves, it is a simple exercise to re-express
this correlator in the basis of cylindrical
multipoles~\cite{Tsang2000}:

\begin{equation*}
\left(\mathbb{D}^0_P\right)_{\alpha^\prime,\alpha}
=\delta_{\alpha^\prime, \alpha} \delta_{s,+}\left(\frac{1-\left|r_{k_\rho,p}\right|^2}{4 q k_\rho} \chi_p
+
\frac{\Im{r_{k_\rho,p}}}{2 |q| k_\rho} \chi_e \right)
\end{equation*}
Here $r_{k_\rho, p}$ are the Fresnel coefficients for a dielectric
plate, $\chi_{p(e)} = 1$ for $k_\rho < \omega$ $(k_\rho > \omega)$ and
zero otherwise, $q = \sqrt{\omega^2 - k_\rho^2}$, and $\delta_{i,j}$
is the Kronecker (Dirac) delta function on discrete (continuous)
indices; the $\delta_{s,+}$ reflects the fact that only waves
propagating in the $+z$ direction are emitted by the plate.  The
expression for the environment correlator $\mathbb{D}^0_E$ is given by
the same expression as $\mathbb{D}^0_P$ with $r = 0$ and
$\delta_{s,+}$ replaced with $\delta_{s,-}$.  Finally, the matrix
elements for the total power flux are
$\left(\mathbb{S}_T\right)_{\alpha^\prime,\alpha} = \frac{2\pi q
  k_\rho}{\omega}\delta_{k_\rho,k_\rho^\prime}
\delta_{p,p^\prime}(-s^\prime)^{\delta_{p^\prime,N}}([\chi_p-\chi_e]s)^{\delta_{p,M}}$.

For the surface meshes, we use approximately 2,500 panels (discretized
surface elements) to get $1\%$ convergence, with the panels highly
concentrated on the area of the objects nearest to the plate.  We
retain angular moments up to $|m| = 10$, and for each $m$ we perform
the $\omega$ and $k_\rho$ integrations using 28 and 48 Gaussian
quadrature points, respectively.  For our study, object $A$ is
composed of doped silicon while the substrate $B$ is silica.  For the
doped silicon dispersion we use a standard Drude-Lorentz
model~\cite{Duraffourg06} with a dopant density of $1.4\times
10^{19}\mathrm{cm}^{-3}$, while for silica we use measured optical
data~\cite{Shen09}.


\emph{Results}: ~\Figref{total_power} shows the geometry-dependence of
the total heat transfer rate between different compact objects and a
dielectric plate, over surface-surface separations $z$ from several
microns down to $20\,\mathrm{nm}$.  In addition to the expected
near-field enhancement, we observe several crossings as, e.g., the
broader surface area of the $R = 0.5\,\mu\mathrm{m}$ radius sphere
competes with the smaller but flatter surface of the $d =
0.4\,\mu\mathrm{m}$ diameter cylinder.  For smaller $z$, the ratio of
the transfer between the $d = 0.4\,\mu\mathrm{m}$ and $d =
0.2\,\mu\mathrm{m}$ cylinders approaches the ratio of their surface
areas (within $6\%$ at $z = 20\,\mathrm{nm}$), as would be expected
from a proximity approximation (PA)~\cite{Narayanaswamy2008b,
  Kruger11}.  The sphere-plate exhibits the $1/z$ power law as
predicted by PA~\cite{Shen09, RosseasuSiria09, Kruger11} to within
$10\%$ for $z < 0.1\,\mu\mathrm{m}$, while the cylinder-plate exhibits
agreement to within approximately $10\%$ over this range using a PA
based on the integral of the plate-plate heat transfer rate over the
cylinder front face and vertical sidewalls.  The contribution from the
sidewalls can be ignored (leading to a $\sim 1/z^2$ transfer
rate~\cite{Volokitin01:PRB}) for $z/d \lesssim 0.01$.  In contrast to
the sphere and cylinders, the cones do not seem to be asymptoting to a
power law, and may even have a \emph{logarithmic} divergence as
$z\rightarrow 0$, a fact which we attribute to the scale-invariance of
the plate-cone configuration when $z \ll 1 \mu\mathrm{m}$ and $z \ll
\hbar c / k_B T$ (the latter eliminating material dispersion effects).
To check the accuracy of our numerical scattering method, we also plot
the results for the sphere where $\mathbb{F}_A$ is calculated
semi-analytically~\cite{Kruger11}, shown as red dots, which agrees to
within $1\%$.

\begin{figure}[tb]
\includegraphics[width=0.45\textwidth]{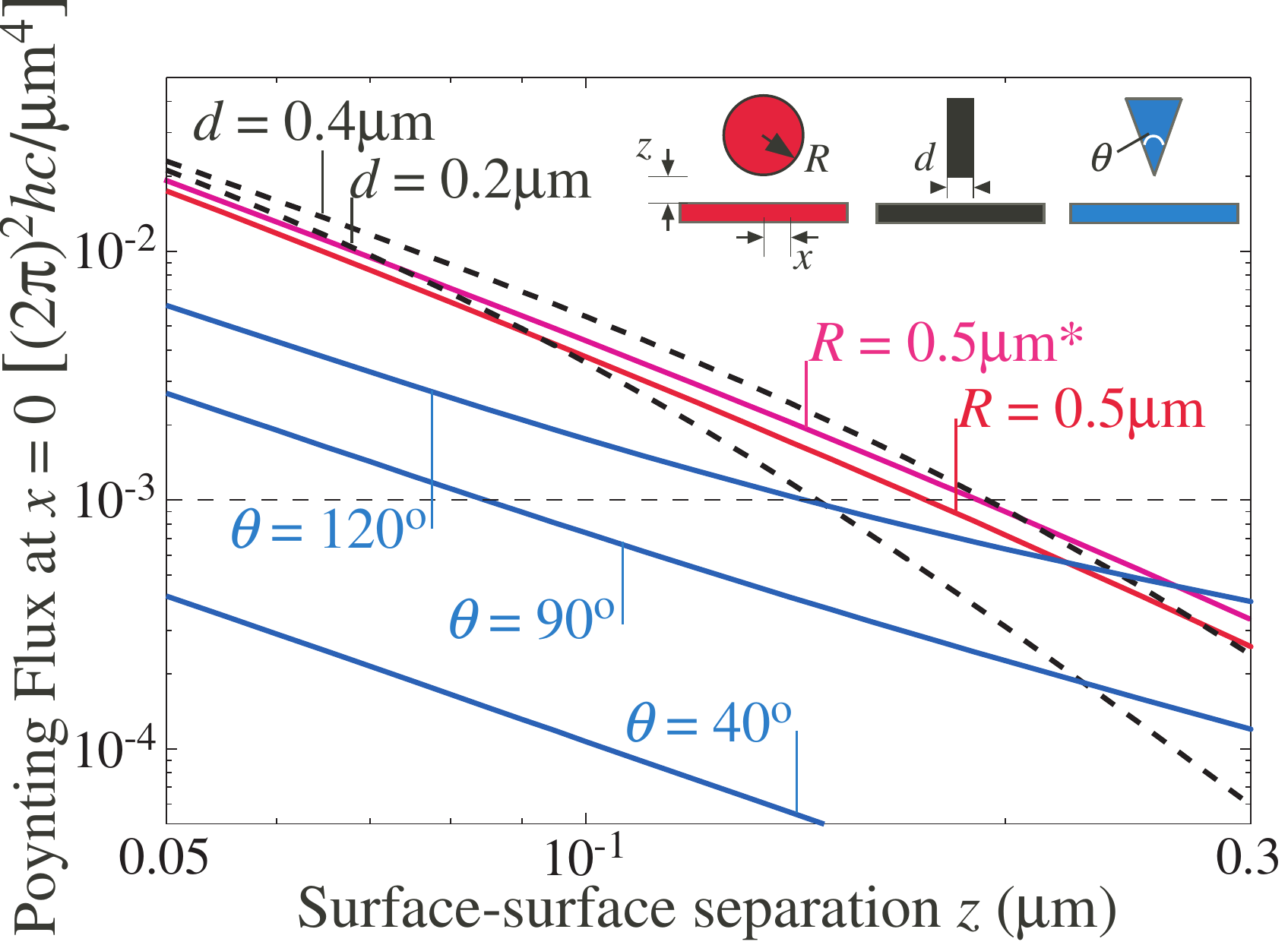}
\centering
\caption{Poynting flux at the origin for the geometries
  of~\figref{total_power} with plate/environment temperature $T_P =
  300\,\mathrm{K}$ and object temperature $T_A = 600\,\mathrm{K}$,
  using the single-polarization approximation (SPA).  The magenta line
  denotes the sphere-plate without the SPA, and the horizontal dashed
  line denotes the threshold used for the crossectional flux profiles
  of~\figref{profiles}.}
\label{fig:peak_poynting}
\end{figure}

For thermal writing applications, an important factor to consider is
not only the total power delivered to the plate, but also the spatial
extent over which this delivery occurs.  In order to examine this, we
envision a scenario in which a critical magnitude of the $z$-directed
Poynting flux is required in order for some change to occur on the
plate, for example, the patterning of a thermal mask for later
etching~\cite{Wilder98}.  \Figref{peak_poynting} plots the Poynting
flux at $x=0$ as a function of $z$, which will tell us how far away
the object must be before it can effect this patterning.  The
cylinders and spheres converge to the same $\sim 1/z^2$ profile for
small $z$ (as expected from a PA), whereas the cones all follow
$1/z^2$ profiles with different coefficients.  This $1/z^2$ dependence
follows from the scale-invariance of the scattering problem for small
$z$, combined with the fact that there is a $1/z$ cutoff in the range
of $k_\rho$ that contributes to the transfer, so that the total number
of modes that contribute is proportional to $\int_0^{1/z} dk_\rho
k_\rho \sim 1/z^2$.  In this calculation we have found that the result
is dominated by the $N$ polarization ($\vec{E} \perp \hat{\vec{z}}$),
mirroring similar phenomena in other near-field cases~\cite{Zhang07},
that results from the behavior of the Fresnel coefficients for high
$k_\rho$.  This is fortunate because we have found that the $M$
contribution to the Poynting flux requires much higher mesh resolution
to converge.  To check this single-polarization approximation (SPA)
for a sphere we also plot the full results, finding that the error
from the SPA is $<20\%$ at the largest $z$, decaying to $<10\%$ at
smaller $z$; SPA for a cone is discussed below.

\begin{figure}[tb]
\includegraphics[width=0.45\textwidth]{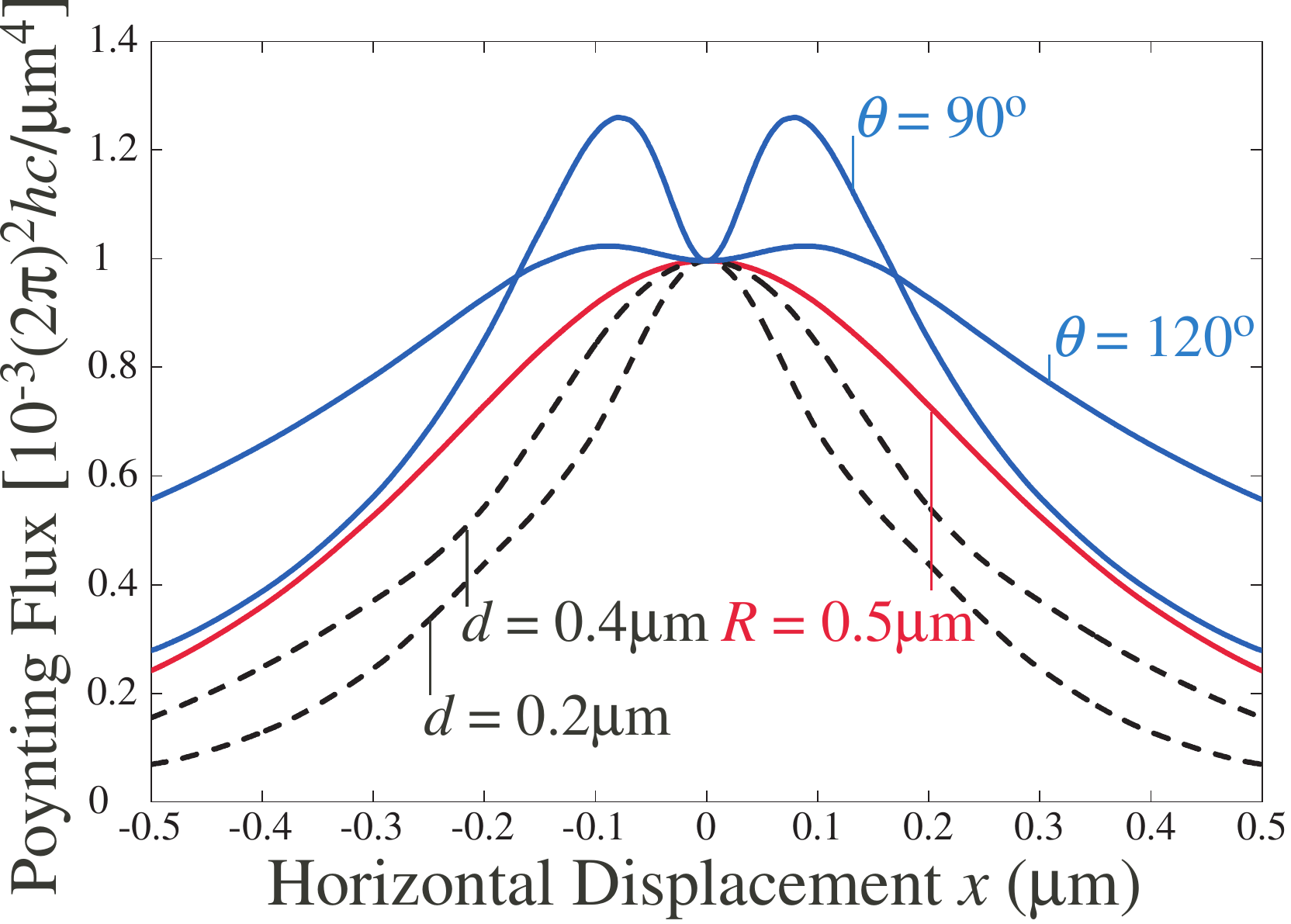}
\centering
\caption{Spatially-resolved heat flux profiles at the substrate
  surface.  $z$ is chosen to fix Poynting flux at $x = 0$ at
  $10^{-3}(2\pi)^2 hc/\mu\mathrm{m}^4$.}
\label{fig:profiles}
\end{figure}

\Figref{profiles} plots the Poynting flux as a function of $x$ showing
the heat transfer profile.  For each object, we chose $z$ to have the
same $x=0$ Poynting flux of $10^{-3} (2\pi)^2 hc/\mu\mathrm{m}^4$
(horizontal dashed line in~\figref{peak_poynting}), corresponding to a
sphere-plate separation of $\approx 200\,\mathrm{nm}$.  The cylinders
and $120^\circ$ cone all reach this threshold at comparable
separations, whereas the $90^\circ$ cone is at less than half the
separation, and the $40^\circ$ cone does not even reach this threshold
within the range considered.

Fixing the peak Poynting flux to $10^{-3} (2\pi)^2
hc/\mu\mathrm{m}^4$, in~\figref{profiles} we plot the Poynting flux
profiles for these shapes as a function of $x$. The widths for the
cylinders are narrower than the sphere, implying that the cylinders
can write higher spatial resolution.  Surprisingly, the cones do not
exhibit this simple behavior.  Rather, the Poynting flux profiles for
the two cones are \emph{non}-monotonic in $x$, with a local minimum at
$x = 0$.  The degree of non-monotonicity appears to increase as the
cone becomes sharper.

\begin{figure}[tb]
\includegraphics[width=0.45\textwidth]{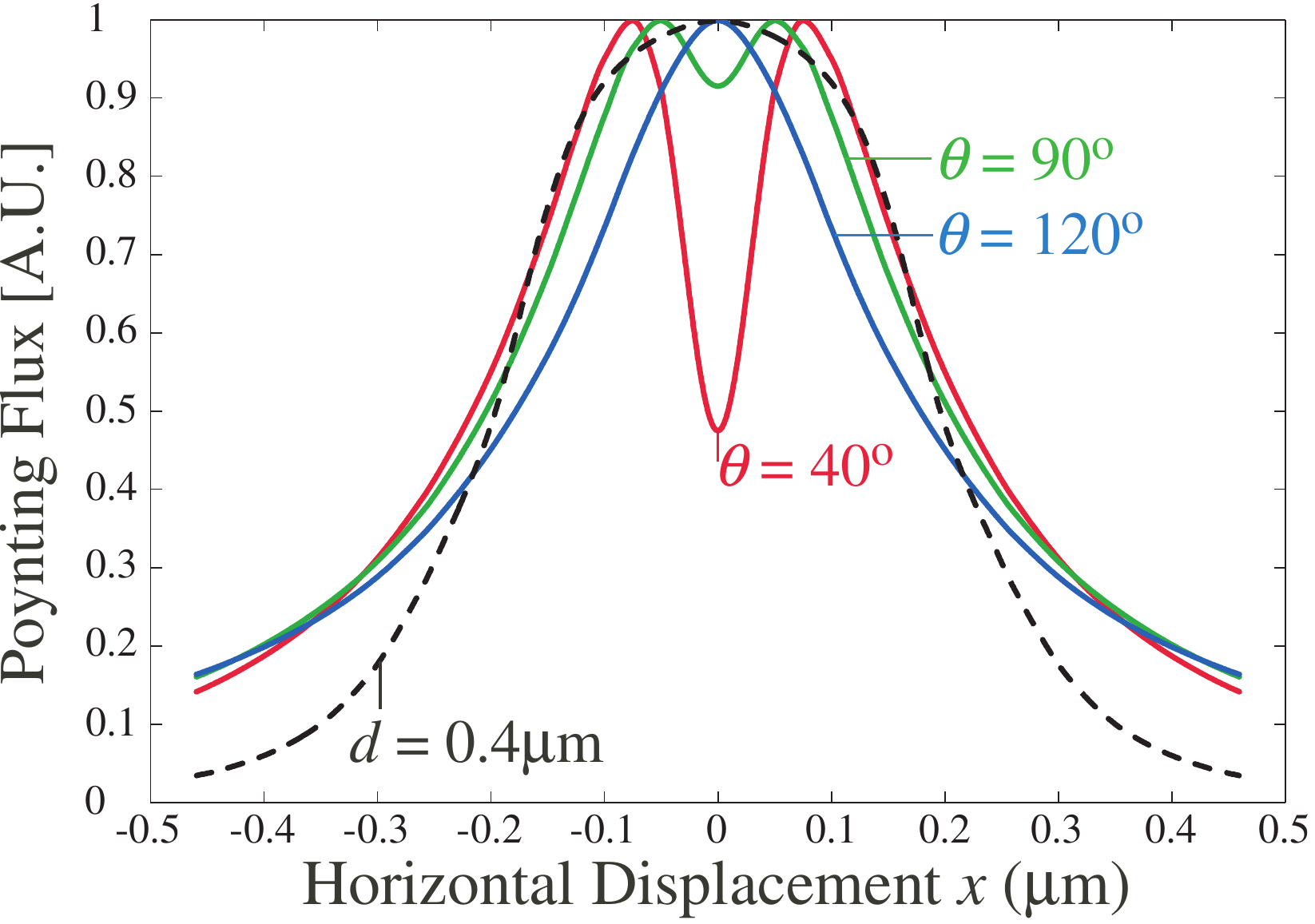}
\centering
\caption{Spatially-resolved heat flux profiles (arbitrary units) at
  the substrate surface for three cones at a single frequency $\omega
  = 0.3066 (2\pi c /\mu\mathrm{m})$ and fixed $z = 70\,\mathrm{nm}$.
  The profiles are normalized so that their maximal value is equal to
  1.  These profiles are computed without the SPA using much finer
  meshes.  For comparison, the profile for a cylinder of radius $d =
  400\,\mathrm{nm}$ (using the SPA) is shown as well.}
\label{fig:full_profiles}
\end{figure}

Before attempting to explain this effect, we must first recall that
the results of~\figref{peak_poynting} and~\figref{profiles} relied on
the SPA; although we know this approximation to work well for flat or
smoothly curved bodies, it is not obvious that it applies equally well
to the cone.  To confirm this result without this approximation, we
must go to a much denser mesh near the cone tip to ensure mesh
convergence; for this, we form a mesh using approximately 12,000
panels for these cones.  We have observed that for the shapes and
separations of interest here, the Poynting flux profiles at all
relevant frequencies have very similar shape, and are simply scaled by
a frequency- and material-dependent weight.  Therefore, it is
sufficient to consider a single frequency, which we pick to be $\omega
= 0.3066 (2\pi c /\mu\mathrm{m})$.  The resulting Poynting flux
profiles for all three cones at a fixed $z = 0.1\,\mu\mathrm{m}$ are
shown in~\figref{full_profiles}; for ease of comparison, all curves
are scaled to have a maximum of 1.  We also show the $d =
0.4\,\mu\mathrm{m}$ cylinder (using the SPA) for comparison.  The dip
at $x = 0$ is less pronounced for the exact curves than for the SPA;
in fact, the dip has vanished for $\theta = 120^\circ$.  However, it
is still present for $\theta = 90^\circ$ and is very prominent for
$\theta = 40^\circ$, where the Poynting flux at $x = 0$ is less than
half of its peak value.  Therefore, we conclude that this effect is
not a result of our approximations.

We believe the explanation for the dip in the Poynting flux is that
as the cone tip becomes sharper, its radiation pattern approaches that
of a dipole with axis normal to the plate, which has zero Poynting
flux at $x =0 $.  This explanation predicts that a very thin cylinder
with $d \ll z$ should also have a dip in the Poynting flux at $x =
0$, which we have also confirmed numerically.

This work was supported by the Army Research Office through the ISN
under Contract W911NF-07-D-0004 and by DARPA under Contract
No. N66001-09-1-2070-DOD and by DFG grant No. KR 3844/1-1.



\end{document}